\documentclass{article}
\usepackage{algorithm2e}
\usepackage{hyperref,mathrsfs,bm}
\usepackage{epsfig,graphics}
\usepackage{color}
\usepackage[applemac]{inputenc}
\usepackage{amssymb,amsmath}
\usepackage{color}
\usepackage{graphicx} 
\usepackage{epstopdf} 
\usepackage{enumerate}
\begin{document}
{\centerline{\large\bf {RIO: A NEW COMPUTATIONAL FRAMEWORK FOR}}
\centerline{\large\bf {ACCURATE INITIAL DATA OF BINARY BLACK HOLES}}}
\vspace*{0.245truein}
\centerline{\footnotesize{\large W. Barreto\footnote{Centro de F\'\i sica Fundamental, Facultad de Ciencias, Universidad de Los Andes, M\'erida, Venezuela}$^{,2}$, P. C. M. Clemente\footnote{Departamento de F\'\i sica Te\'orica, Instituto de F\'\i sica A. D. Tavares, Universidade do Estado do Rio de Janeiro,  R. S\~ao Francisco Xavier, 524, Rio de Janeiro 20550-013, RJ, Brasil}, H. P. de Oliveira$^2$, B. Rodriguez-Mueller\footnote{Computational Science Research Center, San Diego State University, United States of America}}}
\baselineskip=12pt
\vspace*{0.21truein}
\date{\today}
\begin{abstract}

We present a computational framework ({\sc Rio}) in the ADM 3+1 approach for numerical relativity. This work enables us to carry out high resolution calculations for initial data of two arbitrary black holes. We use the transverse conformal treatment, the Bowen-York and the puncture methods. For the numerical solution of the Hamiltonian constraint we use the domain decomposition and the spectral decomposition of Galerkin-Collocation. The nonlinear numerical code solves the set of equations for the spectral modes using the standard Newton-Raphson method, LU decomposition and Gaussian quadratures. We show the convergence of the {\sc Rio} code. This code allows for easy deployment of large calculations. We show how the spin of one of the black holes is manifest in the conformal factor. 

\vspace*{0.21truein}
\noindent Key words: Numerical Relativity; ADM 3+1 formulation; Galerkin-Collocation; Domain decomposition; Binary Black Hole system ; Initial data. 
\end{abstract}

%S-I
\section{Introduction}\label{SI}
% Most recent lectures and advances in the field

Gravitational waves (GWs) from binary black holes (BBHs) were detected by the Laser Interferometer Gravitational-Wave Observatory (LIGO) \cite{abbotetal1}, \cite{abbotetal2}, \cite{abbotetal3}. It is apparent that it will be possible to infer the equation of state and many other features from consolidated observations extended to other laboratories. For instance, with the triangulation LIGO-Virgo \cite{abbotetal4} it was possible to confirm the (tensorial) polarization nature of the gravitational radiation, as predicted by General Relativity. 
These discoveries mark the dawn of gravitational wave astronomy. 
The identification of signals in different bands (gravitational, electromagnetic and neutrinos) was intensified \cite{lehner}. The multi-messenger astronomy era arrives,  offering vast volumes of data from various physical sources (particles, electromagnetic and gravitational radiation) to study a unique cataclysmic event. A large gravitational-wave signal ($\sim$ 100 s; GW170817) was followed by a short gamma-ray burst (GRB 170817A) and an optical transient (SSS17a/AT 2017gfo) found in the host galaxy NGC 4993. The source was detected across the electromagnetic spectrum --in the X-ray, ultraviolet, optical, infrared, and radio bands-- over hours, days, and weeks \cite{abbotetal5}, \cite{abbotetal6}. 
% Link to the next paragraph
There are many open observational issues in which large-scale computations with high resolution are required. 

% Numerical Relativity relevance and alternative/independent catalogs
Numerical General Relativity played a crucial role in the analysis of the detected signals. BBH and BNS simulations were involved in the construction of the phenomenological and effective-one-body waveform models used in the analysis \cite{cl16}. Catalogs of parameters, initial data and waveforms have been published \cite{campanelli}. They contain information on the initial data for the simulations, the waveforms extrapolated to infinity, the maximum luminosity and the physical properties of the remanent black hole. The waveforms are used to interpret the signals detected by the interferometers, and the remanent properties can be used to model the merger of the system from initial configurations. 
It is critical to ensure that the initial data represent --with accuracy-- features of astrophysical scenarios. The correct identification of the initial configuration will help to obtain greater confidence in the analysis of waveforms for the medium and final stages, that is, the coalescence and ringdown. In this context, it is desirable to get initial data which describe with high precision BBH, BNS, and binary black hole-neutron star (BH-NS) systems. These multiple astrophysical scenarios share the same mathematical structure. 
Besides, the most recent discoveries are offering the opportunity to make black hole spectroscopy a reality, possible only theoretically up to 2016 \cite{laguna1}.
It is convenient to report independent catalogs for analysis of the observed data indicating the resolution/precision for each setting. 

% Here we report such a basic an alternative infrastructure.

Here we report a new computational framework based on the Galerkin-Collocation {(GC)} domain decomposition {(DD)} method to solve the Hamiltonian constraint equation for the initial data of Bowen-York puncture black holes. It turns out that the Hamiltonian constraint is a three-dimensional nonlinear elliptic PDE that requires efficient and high precision algorithms for obtaining the appropriate solutions.  Despite the other computational schemes for solving nonlinear elliptic PDEs \cite{thomas}, \cite{NumericalRecipes} we show that the GC-DD method, with its unique aspects, is accurate and computationally economical. Therefore, it leads us to high-resolution solutions with reasonable computational cost. Also, large computing is mandatory for accurate and efficient construction of an initial data catalog, and consequently for the evolution itself.

We have organized the paper as follows. In section \ref{SII} we summarize how the initial data is constructed using the transverse conformal treatment, the Bowen-York, and the puncture methods. We include in this section the ADM mass definition for the sake of completeness. In section \ref{SIII} we present the {GC-DD}  method.  The numerical implementation to get the initial data for an arbitrary BBH is explained in section \ref{SIV}. We focus on the strategy and high-performance computation issues. 
Section \ref{SV} presents the concluding remarks and the outlook for future work.
%%%%%%%%%%%%%%%%%%%%%Figures
\begin{figure}[htb]
\includegraphics[height=7.0cm,width=6.0cm]{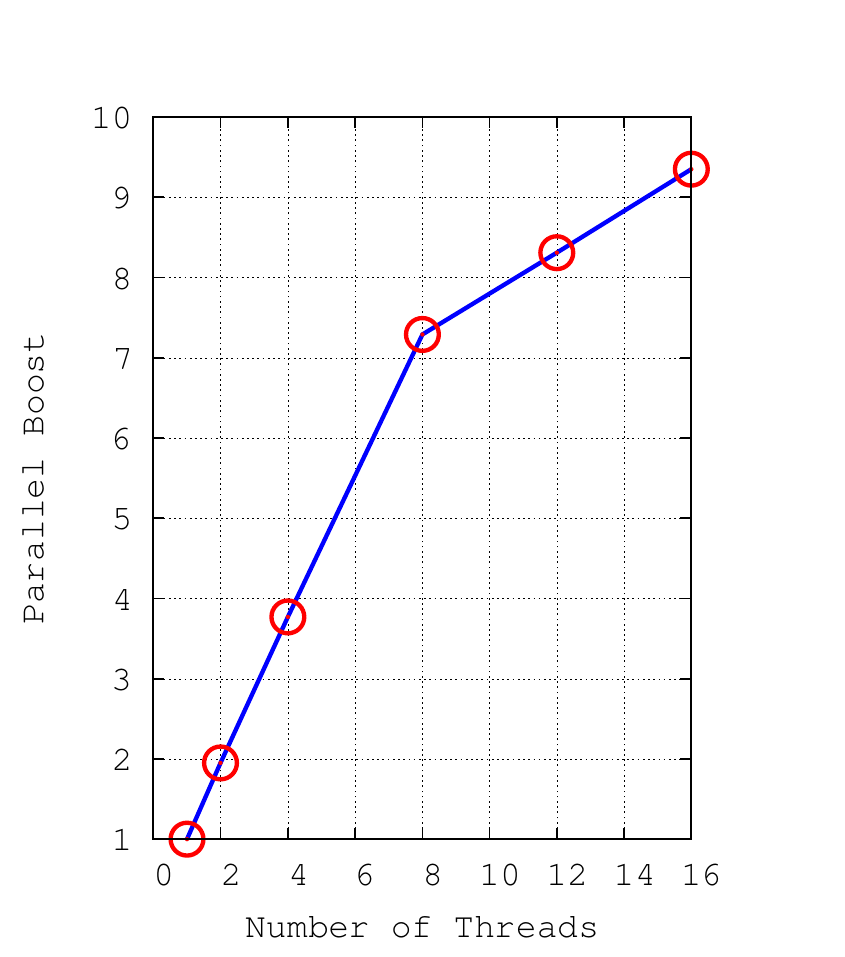}
\includegraphics[height=7.0cm,width=6.5cm]{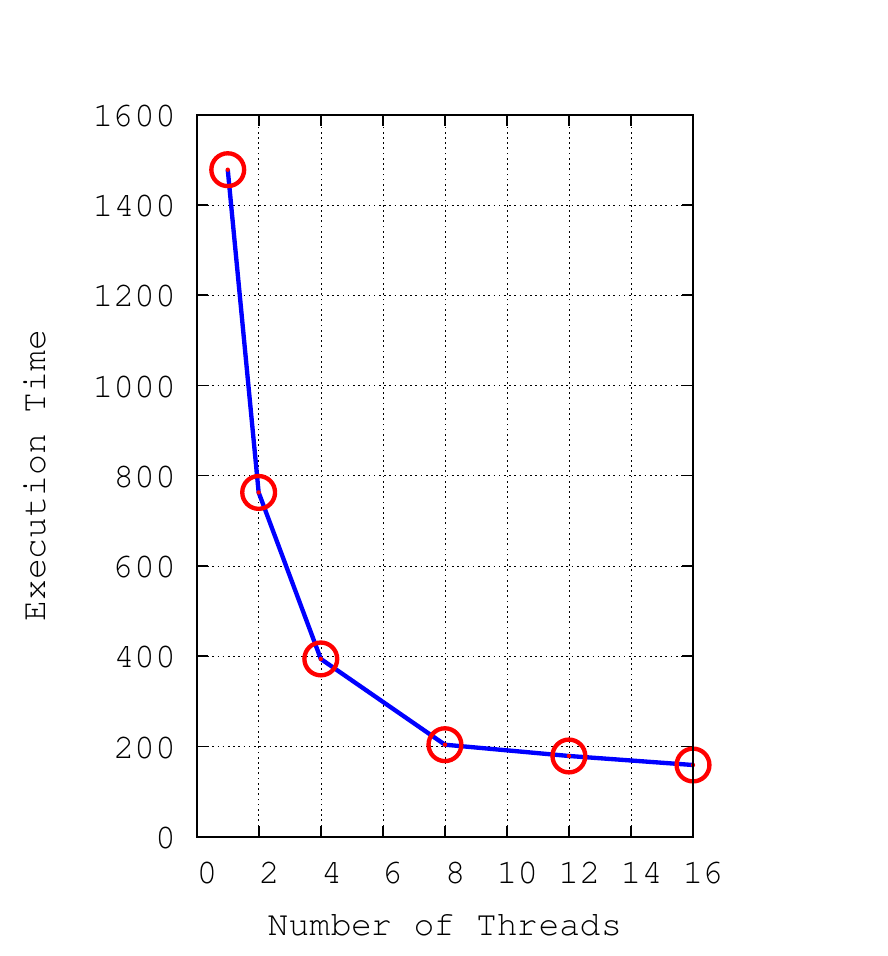}
\caption{Parallel boost factor (left panel) and Execution time (right panel; in seconds) as a function of the number of threads, for one Newton-Raphson iteration and problem size $N_x^{(1)}=N_x^{(2)}=N_y=6$ as reference. The initial trial for $z_n=10^{-3}$, for $n=1...N_z$.}
\end{figure}
\begin{figure}[htb]
\includegraphics[height=6.0cm,width=5.0cm]{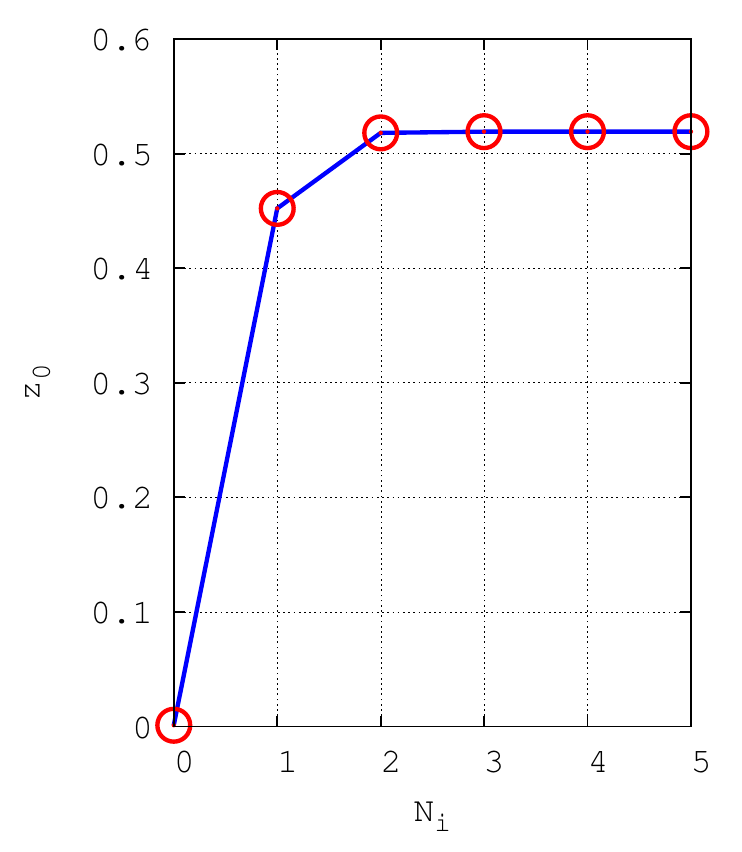}
\includegraphics[height=6.0cm,width=6.0cm]{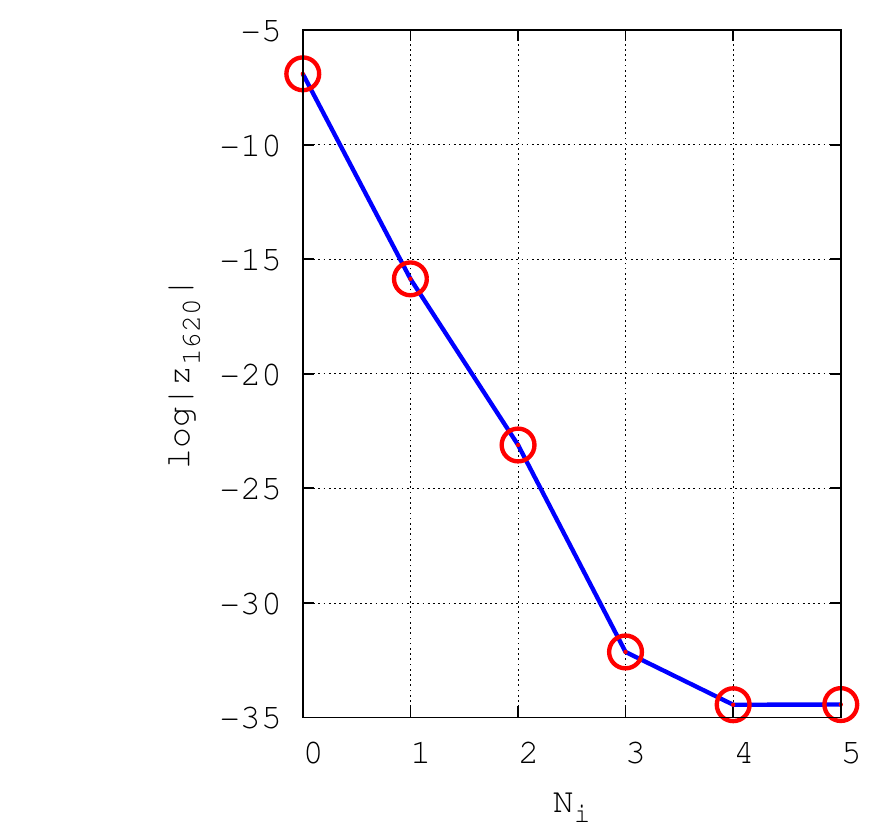}
\caption{Modes $z_0$ (left panel) and $z_{1620}$ (right panel) as a function of the Newton-Raphson iteration $N_i$, for the problem size $N_x^{(1)}=6$, $N_x^{(2)}=12$, $N_y=8$. The physical parameters for the BBH are $m_1=1.5$, $m_2=1$,  $S_0=0.5$, $P_0=2.0$, $s_f=2.0$ 
and $\vec P_1=(P_0,0,0)$, $\vec P_2=(-P_0,0,0)$, $\vec S_1=(-S_0,S_0,0)$ and  $\vec S_2=(0,s_fS_0,s_fS_0)$. The initial trial for $z_n=10^{-3}$, for $n=1...N_z$.}
\end{figure}
\begin{figure}[htb]
\includegraphics[height=7.cm,width=11.cm]{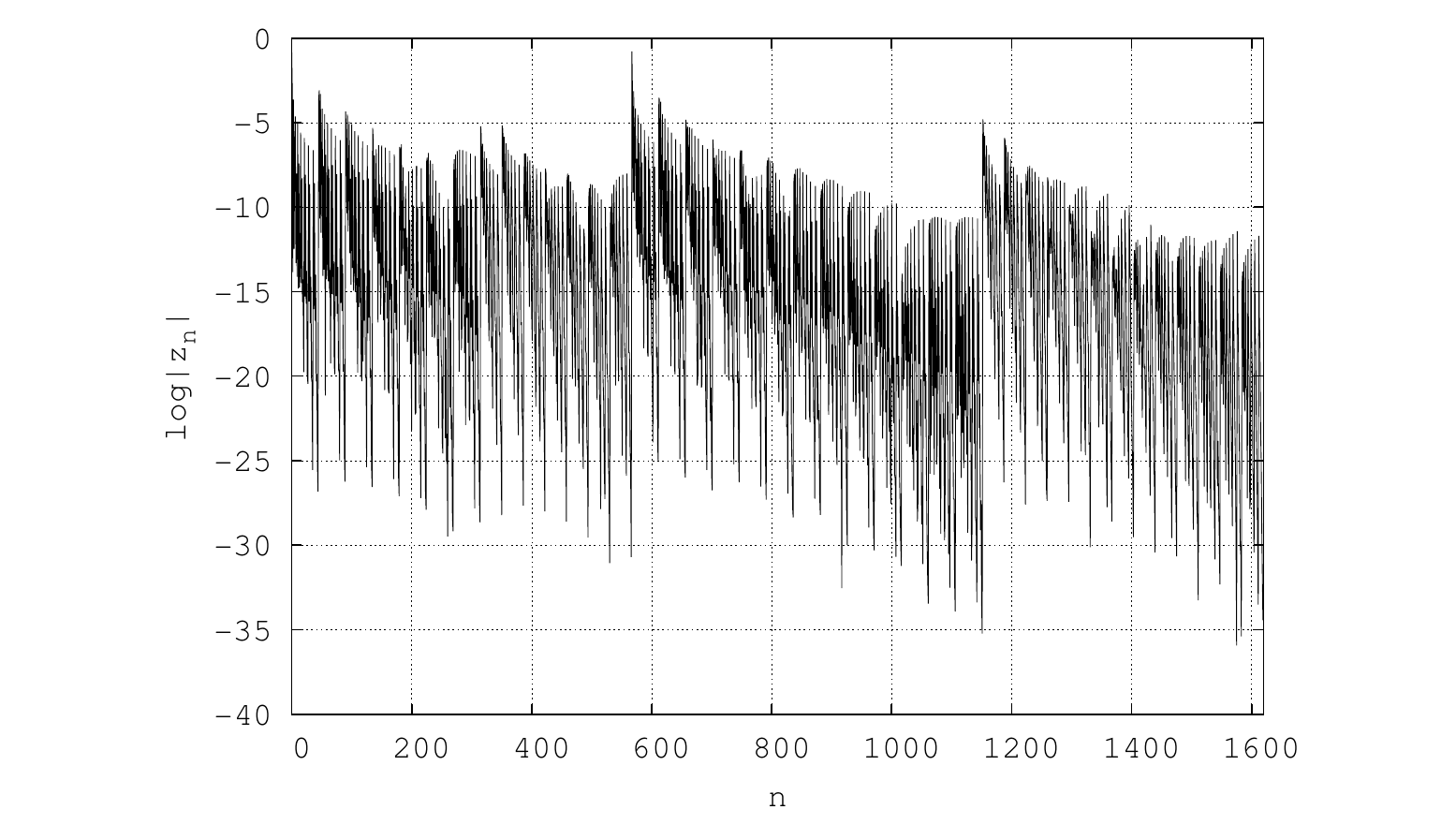}
\caption{Last iterated modes $z_{n}$ as a function of $n$, for the problem size $N_x^{(1)}=6$, $N_x^{(2)}=12$, $N_y=8$. The physical parameters for the BBH are $m_1=1.5$, $m_2=1$,  $S_0=0.5$, $P_0=2.0$, $s_f=2.0$ 
and $\vec P_1=(P_0,0,0)$, $\vec P_2=(-P_0,0,0)$, $\vec S_1=(-S_0,S_0,0)$ and  $\vec S_2=(0,s_fS_0,s_fS_0)$. The initial trial for $z_n=10^{-3}$, for $n=1...N_z$.}
\end{figure}
\begin{figure}[htb]
\begin{center}
\includegraphics[height=7.0cm,width=9.0cm]{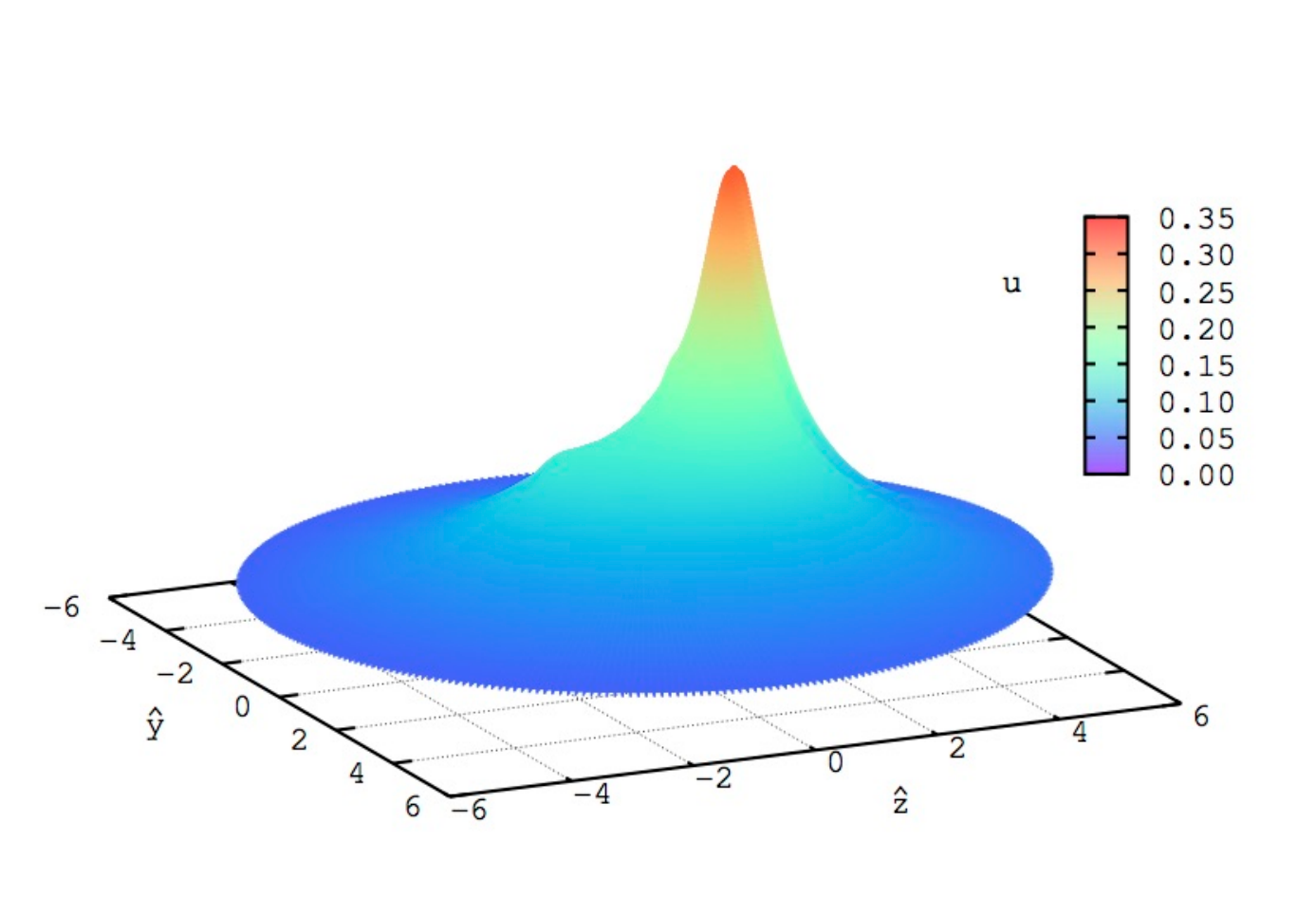}
\end{center}
\caption{Conformal factor $u$ 
as a function of $\hat y$ and $\hat z$, for $\hat x=0$, where the auxiliary Cartesian coordinates $(\hat x,\hat y,\hat z)$ are defined by means the spherical coordinates $(r,\theta,\phi)$ by $\tan\theta=\hat\rho/\hat z$, $\hat\rho=(\hat x^2 + \hat y^2)^{1/2}$, $\tan\phi=\hat y/\hat x$, $r=(\hat\rho + \hat z)^{1/2}$. The set of parameters for this calculation are: $N_x^{(1)}=6$, $N_x^{(2)}=12$, $N_y^{(1)}=8$; $m_1=1.5$, $m_2=1$,  $S_0=0.5$, $P_0=2.0$, $s_f=2.0$ 
and $\vec P_1=(P_0,0,0)$, $\vec P_2=(-P_0,0,0)$, $\vec S_1=(-S_0,S_0,0)$ and  $\vec S_2=(0,s_fS_0,s_fS_0)$. The initial trial for $z_n=10^{-3}$, for $n=1...N_z$.}
\end{figure}
\begin{figure}[htb]
\begin{center}
\includegraphics[height=7.0cm,width=6.cm]{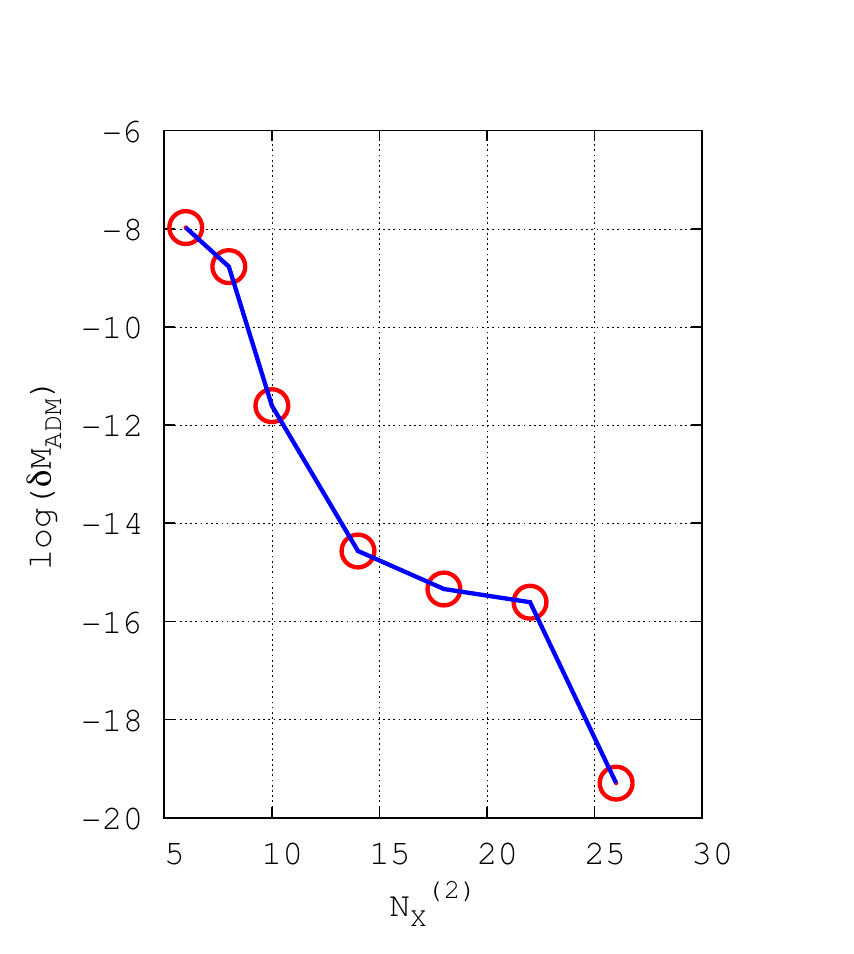}
\end{center}
\caption{Convergence of the ADM mass for a general three-dimensional initial data of a binary spinning-boosted of black holes.}
\end{figure}
\begin{figure}[htb]
\begin{center}
\includegraphics[height=7.0cm,width=6.0cm]{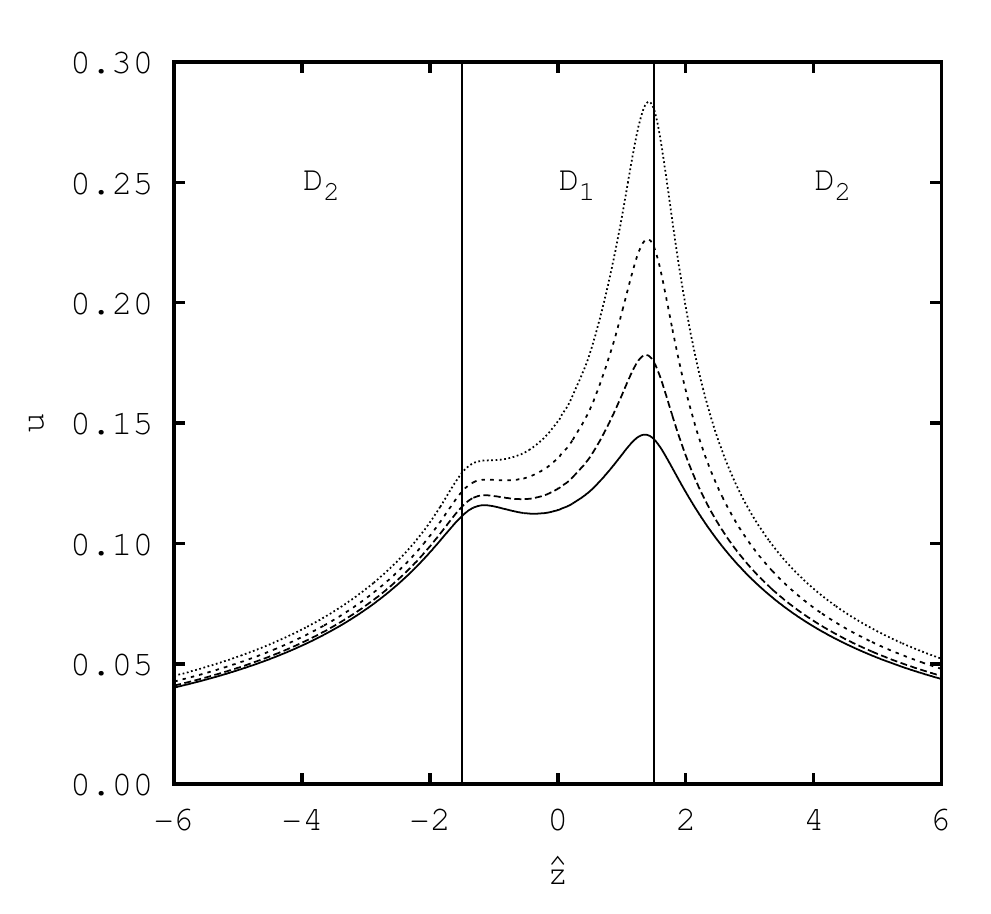}
\end{center}
\caption{Conformal factor as a function of $\hat z$, for the problem size $N_x^{(1)}=10$, $N_x^{(2)}=30$, $N_y=6$. In this case the physical parameters for the BBH are $m_1=1.5$, $m_2=1$,  $S_0=0.5$, $P_0=2.0$  
and $\vec P_1=(P_0,0,0)$, $\vec P_2=(-P_0,0,0)$, $\vec S_1=(-S_0,S_0,0)$ and  $\vec S_2=(0,s_fS_0,s_fS_0)$. Here we consider different values of $s_f$: 0.5 (continuous line); 1.0 (dashed line); 1.5 (short-dashed line); 2.0 (dotted line).}
\end{figure}

%%%%%%%%%%%%%%%%%%%%% Figures 
%S-II
\section{Initial data for an arbitrary system of two black holes}{\label{SII}}
In the ADM 3+1 formulation, the vacuum Hamiltonian and the momentum constraint equations of General Relativity reads
\cite{bs10}:
\begin{equation}
R^2 + K^2 -K_{ij}K^{ij}=0 \label{hc}
\end{equation}
and 
\begin{equation}
\nabla_i(K^{ij}-\gamma{ij})=0, \label{mc}
\end{equation}
where $\gamma_{ij}$ is the 3-metric, $K_{ij}$ the extrinsic curvature of the spacelike hypersurfaces, $K$ its trace, $R$ and $\nabla_i$ the Ricci scalar and the covariant differentiation, respectively, associated with $\gamma_{ij}$.

Following the standard conformal-transverse-traceless decomposition method \cite{bs10}, \cite{y79}, \cite{c00},  we make the following assumptions for the metric and the extrinsic curvature:
\begin{equation}
\gamma_{ij}=\psi^4\bar\gamma_{ij},
\end{equation}
\begin{equation}
K_{ij}=A_{ij}+\frac{1}{3}\gamma_{ij}K,
\end{equation}
\begin{equation}
A_{ij}=\psi^{-2}\bar A_{ij},
\end{equation}
\begin{equation}
\bar A_{ij} = \bar A^{TT}_{ij} + (\bar L {\cal{W}})_{ij} \label{extrinsic},
\end{equation}
with $(\bar L {\cal{W}})^{ij}\equiv \bar\nabla^i{\cal{W}}^j+\bar\nabla^j{\cal{W}}^{i}-\frac{2}{3}\bar\gamma^{ij}\bar\nabla_k{\cal{W}}^k$ and $\bar\nabla_i\bar A^{ij}_{TT}=0$. The operator $\bar\nabla_i$ denotes the covariant differentiation associated with the conformal spatial metric $\bar\gamma_{ij}$. $A_{ij}$ and $\bar A_{ij}$ are traceless and $\bar A^{TT}_{ij}$ is transverse. With these transformations, Eqs. (\ref{hc}) and (\ref{mc}) reduce to
\begin{equation}
8\bar \Delta\psi -\psi\bar R -\frac{2}{3}\psi^5 K^2 + \psi^{-7}\bar A_{ij} \bar A^{ij} = 0, \label{hc_y}
\end{equation}
\begin{equation}
(\bar\Delta_L {\cal W})^i-\frac{2}{3}\psi^6\bar\nabla^i K=0, \label{mc_y}
\end{equation}
respectively, with $(\bar\Delta_L {\cal W})^i\equiv \bar\nabla_j (\bar L {\cal{W}})^{ij}$, $\bar\Delta\equiv \bar\nabla_i \bar\nabla^i$, and $\bar R$ the Ricci scalar of the conformal space. Assuming conformal flatness ($\bar\gamma_{ij}=\eta_{ij}$), maximal slicing ($K=0$) and $\bar A^{TT}_{ij}=0$  the constraint equations (\ref{hc_y}) and (\ref{mc_y}) adopt the form
\begin{equation}
\bar\Delta\psi+\frac{1}{8}\psi^{-7} \bar A_{ij}\bar A^{ij}=0, \label{hc_flat}
\end{equation}
\begin{equation}
(\bar\Delta_L {\cal W})^i=0. \label{mc_flat}
\end{equation}
Construction of the initial data reduces to solve Eq. (\ref{mc_flat}) for ${\cal W}^i$ to get $\bar A^{ij}$ and finally solving Eq. (\ref{hc_flat}) to obtain $\psi$. 

Bowen and York \cite{by80} found that point-source solutions to Eq. (\ref{mc_flat}) are given by 
\begin{equation}
{\cal W}^i=-\frac{1}{4r}[7P^i+ n^iP_jn^j],
\end{equation}
\begin{equation}
{\cal W}^i=\frac{1}{r^2}\epsilon^{ijk}n_jS_k,
\end{equation}
with $\mathbf n=\mathbf r/r$ a unit radial vector. The constant vectors $P^i$ and $S_i$ are interpreted as the linear and angular momentum of the BH. From Eq. (\ref{extrinsic}) the extrinsic curvature associated to these solutions are
\begin{equation}
\bar A^{ij}=\frac{3}{2r^2}[P^in^j+P^jn^i -(\eta^{ij}-n^in^j)P^kn_k], \label{A1}
\end{equation}

\begin{equation}
\bar A^{ij}=\frac{6}{r^3}n^{(i}\epsilon^{j)kl}S_kn_l. \label{A2}
\end{equation}

We consider the puncture method that, allows us to express the conformal factor $\psi$ as,
\begin{equation}
\psi = 1 + \frac{1}{2}\left(\frac{m_1}{|\mathbf r-\mathbf r_1|}+\frac{m_2}{|\mathbf  r-\mathbf r_2|}\right) + u 
\end{equation}
\noindent where $m_a$ is the puncture mass and $|\mathbf{r}-\mathbf r_a|$ denotes the coordinate distance to the center of the black hole located at $\mathbf{r}_a$, $a=1,2$. 

The function $u$ is regular everywhere and satisfies the Robin boundary condition,
\begin{equation}
u = \mathcal{O}(r^{-1}),\label{robin}
\end{equation}
for large distances from the binary system. This condition is a natural consequence of the multipole expansion of the conformal factor.  
As a consequence of the linearity of the momentum constraint, the total background extrinsic curvature corresponding to an arbitrary binary black hole is \begin{eqnarray}
\bar{A}^{ij} = \bar{A}^{ij}_{\mathbf{P}_1} + \bar{A}^{ij}_{\mathbf{S}_1} + \bar{A}^{ij}_{\mathbf{P}_2} + \bar{A}^{ij}_{\mathbf{S}_2}, 
\end{eqnarray}
\noindent where $\bar{A}^{ij}_{\mathbf{P}_a}$ and $\bar{A}^{ij}_{\mathbf{S}_a}$ correspond, respectively, to the background extrinsic curvature of the puncture located at $\mathbf{r}_a$, carrying linear momentum $\mathbf{P}_{a}$ and spin $\mathbf{S}_{a}$, and given by Eqs. (\ref{A1}) and (\ref{A2}). Thus we have
\begin{eqnarray}
\bar{A}^{ij}_{\mathbf{P}_a} &=& \frac{3}{2|\mathbf r - \mathbf r_a|} \left[2P^{(i}_{a} n^{j)}_{a}-(\eta^{ij}-n^i_{a}n^j_{a}) {n}^k_a{P}_{ka}\right], \\
\nonumber \\
\bar{A}^{ij}_{\mathbf{S}_a} &=& \frac{6}{|\mathbf r - \mathbf r_a|^3} n^{(i}_{a}\epsilon^{j)kl}S_{ka} n_{l a},
\end{eqnarray}
where $\mathbf{n}_{a}=(\mathbf{r}-\mathbf r_a)/|\mathbf{r}-\mathbf r_a|$.

The Hamiltonian constraint (\ref{hc_flat}) for the arbitrary binary black holes becomes 
\begin{eqnarray}
&&\bar\Delta u + \frac{1}{8}\bar{A}^{ij}\bar{A}_{ij}\times\nonumber\\
&&\left[1+\frac{1}{2}\left(\frac{m_1}{|\mathbf r-\mathbf r_1|}+\frac{m_2}{|\mathbf r-\mathbf r_2|}\right) + u \right]^{-7} = 0. \label{HC}
\end{eqnarray}
We solve this equation using the Galerkin-collocaton domain decomposition method (see next section). This elliptic equation has the form
\begin{equation}
f(u)=\Delta u + \varrho(u)=0.
\end{equation} 

To complete this section, we introduce the ADM mass for the arbitrary binary black holes \cite{bs10}

\begin{equation}
M_{ADM}=-\frac{1}{2\pi}\int_{\partial \Sigma_\infty} d\bar S_i \bar\nabla^i\psi,
\end{equation}
where $\partial \Sigma_\infty$ is a surface at infinity on the spacelike foliation $\Sigma$; $d\bar S_i$ is an outward surface element.

%S-III
\section{The Galerkin-Collocation two domain method}\label{SIII}

For the sake of completeness, we briefly present the {DD} algorithm based on the {GC} method to generate spacetimes with arbitrary binary black holes (for a general view see \cite{bcdr18} and references therein).
We solve the Hamiltonian constraint (\ref{HC}) by dividing the spatial domain in two domains. We denote these domains by $\mathcal{D}_1: 0 < r \leq r_0$ and $\mathcal{D}_2: r_0 \leq r < \infty$, where $r_0 \geq a$, where $a$ is half of the black holes separation and $(\theta,\phi)$ are the common angular coordinates to both domains.

For the spectral approximation the function $u(r,\theta,\phi)$ at each domain has to satisfy the boundary conditions.  We have established that the function $u$ at each domain is $u^{(A)}(r,\theta,\phi)$, $A=1,2$ accounting for the domains $\mathcal{D}_1,\mathcal{D}_2$, respectively. The spectral expansions in each domain are expressed as

\begin{equation}
u^{(A)} = \sum^{N^{(A)}_x,N_y}_{k,l = 0}\sum^{l}_{m=-l}\,c^{(A)}_{klm} ~\chi^{(A)}_{k}(r) Y_{lm}(\theta,\phi). \label{SE}
\end{equation}

\noindent Here $c^{(A)}_{klm}$ represents the unknown coefficients or modes, $N^{(A)}_x$ and $N_y$ are, respectively, the radial and angular truncation orders; the spherical harmonics, $Y_{lm}(\theta,\phi)$, are the basis functions for the angular patch. 

The radial basis functions, $\chi^{(A)}_{k}(r)$, follow the prescription of the Galerkin method in which each basis function satisfies automatically the boundary conditions. They are expressed by combinations of the Chebyshev polynomials along with the introduction of convenient mappings adapted to each domain.

The entire radial domain $0 \leq r < \infty$ is mapped onto the interval $-1 \leq x < 1$ through the algebraic map \cite{boyd},
\begin{equation}
r = L_0\frac{(1+x)}{1-x},
\end{equation}
where $L_0$ is the map parameter. The domains $\mathcal{D}_1$ and $\mathcal{D}_2$ are characterized by $-1 \leq x \leq x_0$ and $x_0 \leq x < 1$, respectively, and $x_0$ is related to $r_0$ by $r_0 = L_0(1+x_0)/(1-x_0)$. We further define linear transformations $x^{(A)}=x^{(A)}(x)$, such that $-1 \leq x^{(A)} \leq 1$. The collocation points are designated by $x^{(A)}_k$ and mapped back to $r_k$ in the radial physical domain. 

Obviously the coefficients $c^{(A)}_{klm}$ must be complex. Since the conformal factor is a real function, the real and imaginary parts of $c^{(A)}_{klm}$ satisfy the following symmetry relations
\begin{equation}
c^{(A)*}_{kl-m}=(-1)^{-m}\,c^{(A)}_{klm},
\end{equation}
\noindent due to $Y^*_{l-m}(\theta,\phi)=(-1)^{-m}Y_{lm}(\theta,\phi)$. Consequently, the number of independent coefficients in each domain is $\left(N^{(A)}_x+1\right)\left(N_y+1\right)^2$. 

We have to guarantee that the spectral approximations of $u^{(1)}$ and $u^{(2)}$ given by Eq. (\ref{SE}) are part of the same function. Thus, we have to impose the appropriate matching conditions at the boundary $r=r_0$ which separates both domains, that is,
\begin{eqnarray}
&&\;\;\;\;\;\;\;[u^{(1)}-u^{(2)}]_{r=r_0}=0, \nonumber \\ \nonumber \\
&&\left[{\frac{\partial u}{\partial r}}^{(1)}-{\frac{\partial u}{\partial r}}^{(2)}\right]_{r=r_0}=0. \label{MC}
\end{eqnarray}
\noindent These relations restrict the total number of independent coefficients.

We obtain the coefficients $c^{(A)}_{klm}$ extending straightforwardly the implementation for one domain \cite{cd16}. From the method of weighted residuals \cite{finlayson}, these coefficients are evaluated with the condition of forcing the residual equation to be zero on average. 

By taking into account the matching conditions, we have $(N_x^{(1)}+N_x^{(2)})(N_y+1)^2$ coefficients to determine. 
Thus, we need $N^{(A)}_x$
radial collocation points in each domain which are calculated 
from the Chebyshev-Gauss collocation points in the computational domain
\begin{eqnarray}
x^{(A)}_j=\cos\left(\frac{j \pi}{N_x^{(A)}+2}\right),
\end{eqnarray}
and 
\begin{eqnarray}
r^{(1)}_j &=& \frac{L_0 (1+x^{(A)}_j)}{2L_0/r_0 + 1-x^{(A)}_j}, \\
\nonumber \\
r^{(2)}_j &=& \frac{2r_0+L_0(1+x^{(A)}_j)}{1-x^{(A)}_j},
\end{eqnarray}
with $j=1,2..,N_x^{(A)}$.
We have excluded the points at infinity since the residual equation is automatically satisfied asymptotically due to the choice of the radial basis function.
Notice that the origin is also excluded.

We can obtain $(N^{(1)}_x+N^{(2)}_x)(N_y+1)^2$ algebraic non linear equations from Eq. (\ref{HC}), which together with the $2(N_y+1)^2$ equations from the matching conditions (\ref{MC}) constitute the set of equations to be solved for the modes
$c^{(A)}_{klm}.$

%S-IV
\section{Numerical implementation and testing}\label{SIV}
The set of equations is written as one vector $H_n(u)=0$, with $n=1...N_z$, and $N_z=(N_x^{(1)} + N_x^{(2)}+2)(N_y+1)^2$. The system $H_n$ has $z_n$ real solutions; each solution $z_n$ corresponds to one and only one real (or imaginary) part of the coefficients $c^{(A)}_{ijk}$. {The way in which $z_n$ is ordered and related with $c^{(A)}_{ijk}$ is displayed in the following pseudo-code:}

\begin{algorithm}[H]
{
 \KwData{$z_n$}
 \KwResult{$c^{(A)}_{ijk}$}
 $n=1$\;
 \For{$A:=1$ \mbox{\bf to} $2$}{
 \For{$i:=0$ \mbox{\bf to} $N_x^{(A)}$}{
 \For{$j:=0$ \mbox{\bf to} $N_y$}{
 \For{$k:=0$ \mbox{\bf to} $j$}{
 $\Re\{c^{(A)}_{ijk}\}:=z_n$; \\
 $n=n+1$;}}}
 \For{$i:=0$ \mbox{\bf to} $N_x^{(A)}$}{
 \For{$j:=0$ \mbox{\bf to} $N_y$}{
 \For{$k:=0$ \mbox{\bf to} $j$}{
 $\Im\{c^{(A)}_{ijk}\}:=z_n$; \\
 $n=n+1$;}}}
 }
 }
\end{algorithm}

{To solve the system of equations we use the standard Newton-Raphson method \cite{NumericalRecipes},
\begin{equation}
J\delta z_n = - H_n,
\end{equation}
where $J=(\partial H_n/\partial z_m)$ is the Jacobian matrix and $\delta z_n$ is the variation of the solution $z_n$ between the iteration $N_i$ and $N_{i-1}$, up to some
specified tolerance for the convergence. 
To get the set of solutions at each iteration, we use an LU or QR decomposition \cite{NumericalRecipes}. We observe the best performance for the LU decomposition.
All the special functions and its derivatives are calculated using the standard generating formulae from the Numerical Recipes library of subroutines.}

The general strategy to solve the problem has the following three main steps:
\begin{enumerate}
\item Input: 
\begin{enumerate}
  \item Set of physical parameters which represent an arbitrary BBH: masses $m_1$ and $m_2$, linear momentum $P_0$ and angular momentum $S_0$, separation between the black holes $a$;
  \item  Set of computational parameters: truncation numbers for both domains $N_x^{(1)}$ and $N_x^{(2)}$, truncation number for the angular domain $N_y$, domain decomposition map length $L_0$, matching point $x_0$; number of iterations and tolerances for the Newton-Raphson method: $N_i$, $\tau_x$ and $\tau_f$; 
  \item A previous solution or start iteratively {\it ab initio} $z_n$;
\end{enumerate}
\item Newton-Raphson procedure:
\begin{enumerate}
\item Obtention of the system of nonlinear equations $H_n$ and the Jacobian;
\begin{enumerate}
     \item Construction of the complex coefficients $c_{ijk}^{(A)}$ from the real $z_n$;
     \item Conformal factor and derivatives;
     \item $H_n$;
     \item Jacobian; for a parallel code the load distribution is performed only here;  
\end{enumerate}
\item Solve the system of equations using LU decomposition;
\item If the conditions selected for the Newton-Raphson procedure are satisfied then the next step is performed. Otherwise, we repeat this main step from (a);

\end{enumerate}
\item Output: 
\begin{enumerate}
\item Last iteration of $z_n$;
\item Conformal factor;
\item ADM mass.
\end{enumerate}    
\end{enumerate}

{The computational framework was implemented  initially in Maple. The Maple script was used to develop a serial code in Fortran and was used as a reference for the numerical validation as well. In cases in which the memory and the velocity were a real limit for simulations, we use only the numerical code in Fortran which is very economic handling memory resources. In turn the Fortran solver let us to identify the sector where the computational cost was located in the problem, to implement a parallelization in the most straightforward way. We want to stress that in practice the algorithms in both cases (for Maple and Fortran) are different although the final result is the same. In fact, one essential difference is the quasi numerical character of the Maple script, which let us validate the Fortran code.}

The parallel code implemented with OpenMP scales linearly into the eight processors range for an AMD Ryzen 7 1800X Eight core processor, depending on hyperthreading. We define the parallel boost factor as the serial code time over the parallel code time for $N$ threads. Considering the execution time we observe that it diminishes a half when we duplicate the number of threads, up to the limit (in this case 8) where the hyperthreading begins. Figure 1 illustrates this performance for a reference size problem. 

We select for testing a general three-dimensional initial data of a binary spinning-boosted of black holes (taken from Br\"ugmann \cite{b99}) with the following features: $m_1=1.5$, $m_2=1$, $S_0=0.5$, $P_0=2.0$, $s_f=2.0$, $\vec P_1=(P_0,0,0)$, $\vec P_2=(-P_0,0,0)$, $\vec S_1=(-S_0,S_0,0)$, $\vec S_2=(0,s_fS_0,s_fS_0)$, with $a=L_0=1.5$ and $x_0=0$. Note that we are using an extra parameter $s_f$ to change the spin for the black hole labeled as 2. For these parameters we solve the system with the following truncation parameters: $N_x^{(1)}=6$, $N_x^{(2)}=12$, $N_y=8$, giving the total number of modes $N_z=1,620$. The initial trial for $z_n=10^{-3}$, for $n=1...N_z$. Figure 2 shows the first and the last mode convergence with the Newton-Raphson iterations. Figure 3 displays the set of all modes $z_{n}$ as a function of $n$ for the last iteration. {Note 
how using the Newton-Raphson method the last iteration of each spectral mode is bounded at each
collocation point.} The apparent randomly behavior is disappears  when we visualize the conformal factor for the BBH setting (see figure 4). Figure 5 shows the convergence of the ADM mass calculated using $\delta M_{ADM}=|1-M_n/M_m|,$ where $M_n$ represents the ADM mass calculated for different truncations in the second domain and $M_m$ the ADM mass calculated for the biggest truncation used in that domain ($M_{30}$).

%\section{Performance for a sample model}
\section{Diminishing the spin for the second black hole}

%[Explain the model]
The main feature of the initial data considered in the last section is its general three-dimensional character. For a BBH with different masses and different linear momenta we explore how the initial spin for one of the black holes can be distinguished in the conformal factor.
%[Selection of size and paramenters]
Considering the size problem $N_x^{(1)}=10$, $N_x^{(2)}=30$, $N_y=6$, we select $s_f=0.5$; $1.0$; $1.5$; $2.0$. The figure 6 shows the conformal factor profile as a function of $\hat z$ for these values of $s_f$. These results suggest a tendency to realize indistinguishable initial black holes as a system, being so different individually. 
%[Report the performance]
Each calculated profile (for each $s_f$) takes 36 hours using 8 threads on a dedicated Intel(R) Core(TM) i7 CPU 920  @ 2.67GHz. 
%S-V
\section{Concluding remarks and outlook of future work}\label{SV}
{We report here a {novel} computational framework which is the implementation of a DD using the GC method, in the context of ADM 3+1 approach in numerical relativity, {to solve the Hamiltonian constraint}. The developed numerical code enables us to perform high-resolution calculations for initial sets of two arbitrary black holes. The numerical code for the framework was tested and named {\sc Rio}. {The RIO code constructs the puncture initial data for two black holes, which is a problem solved by several authors using different numerical methods (see Ref. \cite{c00} and references therein),  \cite{abt04}. In this sense the Einstein Toolkit \cite{ETK} is a good example in the open source vein, which is gaining users, adaptability, technologies
and languages. Nevertheless we think the GC-DD implemented in the RIO code
is an alternative. The GC-DD method, with its unique
aspects, is accurate and computationally economical (using small quantity of memory) and can be
massively implemented.}
{Such a computational procedure is certainly of high computational
cost, but at the same time is highly accurate. It could be useful
as an alternative procedure to disclose details when the initial
data have a complex structure in three spatial dimensions.
Our code is the implementation of a framework which
has been used in other context of Numerical General Relativity \cite{bdr18}.
We report a computational framework which could be straightforwardly extended for fixed mesh refinement.  Methods as Newton-Raphson and LU decomposition, can be easily replaced to gain efficiency.}

Being the highest computational cost in the obtention of the Jacobian for the Newton-Raphson method implementation, it was solved efficiently with a parallel code, which scale strongly (keeping the same size problem). An interested reader in these computational issues can ask about the weak scale (keeping the same load per processor) for the parallel code. Such a characterization is possible but it depends on what subgrid we require high resolution. For example, for the reference size problem $N_x^{(1)}=N_x^{(2)}=N_y=6$, for the second domain we can keep the same load per processor increasing the truncation from the base as $N_x^{(2)}=6,20,34,48,62,...$ increasing the number of processors in $N_p=1,4,9,16,25,...$ respectively. We would like to evaluate the performance using hundreds or thousands of threads (using CPUs and GPUs) to go ahead with a most general code which will lead us to evolve an initial data with high resolution. We are currently developing a more natural and transparent serial/parallel code using  Python modules. Also, we are exploring other alternative numerical procedures to solve the Hamiltonian constraint, such as the Singular Value Decomposition and Broyden methods. 
Additional work is being advanced and will be reported elsewhere.}
%\begin{acknowledgements} 
\section*{Acknowledgments}
The authors acknowledge the financial support of the Brazilian agencies Conselho Nacional de Desenvolvimento Cient\'ifico e Tecnol\'{o}gico (CNPq). H. P. O. thanks Funda\c c\~ao Carlos Chagas Filho de Amparo \`{a} Pesquisa do Estado do Rio de Janeiro (FAPERJ) for support within the Grant No. E-26/202.998/518 2016 Bolsas de Bancada de Projetos (BBP).
%\end{acknowledgements}

\end{document}